\documentclass[aip,jmp,numerical,preprint,superscriptaddress]{revtex4-2}

\usepackage{amsfonts,amsmath}
\usepackage{dcolumn}
\usepackage{bm}
\usepackage{graphicx}
\usepackage{cancel}

\begin{document}

\title {Gauge-invariant perturbation expansion in powers of  electric charge for the
  density-of-states of a network model for charged-particle motion in a
  uniform background magnetic flux density}
\author{F. D. M. Haldane}
\affiliation{Department of Physics, Princeton University \\
Princeton, New Jersey 08544-0708, USA}
\date{November 11, 2020}
\begin{abstract}
 An explicitly-gauge-invariant expansion in powers of $e/\hbar$ times
 the magnetic flux density is formally obtained for the density of states (as
 characterized by  the trace of the resolvent  $\widehat G$ = $(\omega - \hat
 h)^{-1}$) of a  charged particle moving on a Hermitian quantum
 network that is embedded in a  Euclidean background that supports a
 uniform magnetic flux density. The explicit expressions, given here up
 to third order in the flux density, are  also  valid for 
the ``local trace'' (the trace of $\widehat P_i \widehat G$, where
$\widehat P_i$ is the projector on a network node), and do not appear
to have been previously given.   
\end{abstract}

\maketitle

This paper addresses the formal problem of the  gauge-invariant
expansion, in powers of $e/\hbar$ times the magnetic flux density, of
the density of states or spectrum of a quantum network on which a particle tunnels
or ``hops'' between orbitals located at the network nodes,
when the particle is given an electric
change $e$, and the network is embedded in a Euclidean background
that supports a uniform magnetic flux density that ``dresses''  the
amplitude of a closed hopping path on the network with a geometric
Bohm-Ahanorov phase that depends on the embedding of the network in
the background.   (The phase is  for Euclidean straight-line geodesic
tunneling paths between embedding positions of network nodes.)
Surprisingly, the explicitly-gauge-invariant
result given here does not appear to have appeared previously, at least to the
knowledge of  the author, and is required for a modern discussion
(in terms of Berry geometry) of Landau diamagnetism of Bloch states,
which will be will be reported elsewhere.   However, the results and
derivation reported here apply more generally, without requiring Bloch
periodicity or a thermodynamic limit,
and may have other applications, for example to
disordered systems,  so it is useful to present them as a  separate work.

While a continuum Schr\"{o}dinger equation with a spatially-varying
potential was the original microscopic description of electronic
motion  in solid
materials, both spatially periodic and non-periodic, the more common modern
approach takes account of atomic structure, and views the problem of
electron motion inside rigid solid matter
as  finite-range tunneling  (or ``hopping'') between atomic-like
localized orbitals of various point-symmetries
centered at positions defined by the atomic nuclei (these are the
``Wannier orbitals'' if the system is periodic and its electronic
structure has been approximated by
a single isolated Bloch band).   

The restriction of the electronic single-particle Hilbert space to a 
finite density of localized orbitals is a ``renormalization'' procedure
that thins out the electronic spatial degrees of freedom and
replaces continuous electronic motion in Euclidean space by
hopping  motion on a quantum network of  orthogonal localized spin-1/2 orbitals
$|i\sigma\rangle$ defined at network nodes, with links between nodes
(and onsite terms at nodes) provided by the
matrix elements
\begin{equation}
  \langle i\sigma | \hat h |j\sigma'\rangle
  = \sum_{k=0}^3h^{(k)}_{ij} \sigma^k_{\sigma\sigma'} ,
\end{equation}
where $\{\bm \sigma^k ,k = 1,2,3\}$ are the $2\times 2$ unitary
Hermitian Pauli
matrices, and $\bm \sigma^0$   is the $2\times 2$ identity matrix.
In general the matrices  $\bm h^{(k)}$ are complex Hermitian, $(h^{(k)}_{ij})^*$ =
$h^{(k)}_{ji}$,
but if there is
antiunitary time-reversal symmetry, $\bm h^{(0)}$ can generically be
chosen
real symmetric,
and, for $k \in \{1,2,3\}$, $\bm h^{(k)}$ can be chosen  imaginary antisymmetric;
the eigenvalue spectrum then has Kramers degeneracy.
The network description makes the inclusion of spin-orbit coupling
easier, and is the primary tool for describing
topologically-non-trivial band structures.

The free-fermion thermodynamics of the quantum network can be entirely
expressed in terms of the quantity
\begin{equation}
  \textrm{Tr} \left (\widehat G(\omega )\right )
   = \sum_i\textrm{Tr}_i \left (\widehat G(\omega)\right ) \equiv \sum_{i\sigma}\langle
  i\sigma | (\omega -
  \hat h)^{-1}|i\sigma\rangle ,
\end{equation}
the (extensive) trace of the single-particle Green's function,  which
is   expressible as a sum of local contributions from
each network node, using the orthogonal projectors
\begin{equation}
      \widehat P_i \equiv \sum_{i\sigma}|i\sigma\rangle\langle
                     i\sigma| ,\quad \sum_i\widehat P_i = \openone, \quad
                     \widehat P_i  \widehat P_j =  \delta_{ij}\widehat
                     P_i,
\end{equation}
which define the ``local trace'' $\textrm{Tr}_i(\widehat O)$ $\equiv$
$\textrm {Tr}(\widehat P_i\widehat O)$ of an operator $\widehat O$.
(All results given here for the trace  apply equally to the ``local
trace'', which greatly simplifies taking the thermodynamic limit.)
Note that the trace (and the local trace) has a ``locator'' expansion as a weighted  sum over closed paths
on the network.
The free-fermion free energy (grand canonical
thermodynamic
potential)  $\Omega (T,\mu) $  at
fixed temperature $T$ and chemical potential $\mu$, with  $\beta$ = $1/k_BT$,  
is given in terms of $\textrm{Tr}(\widehat G(\omega) ) $  by
\begin{equation}
  -\beta\Omega(T,\mu)= \int _C\frac{d\omega}{2\pi i} \textrm{Tr}\left
  (\widehat G(\omega)\right ) \ln ( 1 + e^{(\beta(\mu
-\omega))}).
\end{equation}
Generally, the spectrum of real eigenvalues $\varepsilon_{\nu}$  of  a Hermitian
network Hamiltonian has both a lower and an upper bound,
$E_{\textrm{ min}} < \varepsilon_{\nu} < E_{\textrm{ max}}$, that remain finite in
the large-network limit if the number of network neighbors that a node is
connected is finite.
Then  (at finite
$\beta$) the contour $C$ in the complex $\omega$-plane can be taken as a closed
contour that encloses all the poles of $\textrm{Tr}(\widehat G(\omega))$
at $\omega$ = $\varepsilon_{\nu}$,
which lie on the real axis in the range $[E_{\textrm{ min}},E_{\textrm{ max}}]$, but none of the
poles of the integrand at $ \omega$ =
$\mu  \pm i (2n + 1)\pi \beta^{-1}/2$, $n$  = $0,1,\ldots , \infty$.

So far the network is an abstract network, and its thermodynamics
is entirely a property of the spectrum of the Hermitian single-particle
operator
$\hat h$, contained in  $\textrm{Tr}(\widehat  G(\omega))$.
What has been lost in restricting the electron dynamics to a network is
any notion of how that network is embedded in a background Euclidean
space.   Specification of the details of the embedding only  becomes necessary when the
fermions are given an electric charge $e$ that allows them to interact
with background electromagnetic fields as they tunnel between nodes on the network.
Unlike the electron states which are now 
localized on the network nodes, the
electromagnetic fields are supported throughout the background
continuous Euclidean space.

The embedding is a key ingredient of
Bloch's theorem that  (provided there is no uniform component of the
electric or magnetic field if the particle carries electric charge) the eigenstates of the one-body Hamiltonian of
a periodic system, which  are quasiperiodic, can be
factorized into a quasiperiodic unitary factor (that depends on the embedding) and a
periodic factor.  In the continuum Schr\"{o}dinger description,
Bloch's theorem seems so ``obvious'' that it is not immediately
apparent that it involves an embedding.    This becomes clearer in the
the case of network or ``tight-binding'' models.   However, it is
often overlooked there too, as calculations of Bloch energy bands (and  their
topological invariants)  do not require specification of the embedding,
and for those calculations, a common practice is to implicitly embed
all the orbitals in the unit cell at its nominal origin, which conveniently
makes the wavefunctions periodic (as opposed to quasiperiodic)
functions of the Bloch vector in the Brillouin zone.     In contrast, the Berry curvatures
of Bloch bands, which  are part of the semiclassical description of the dynamics of
wavepackets in the presence of quasi-uniform electromagnetic fields,
are fundamentally embedding-dependent quantities.

The embedding will formally be in $d$-dimensional Euclidean space,
parametrized as
\begin{equation}
  \bm x = x^a\bm e_a, \quad \bm e_a \cdot \bm e_b = \delta_{ab}.
\end{equation}
The results will in fact be independent of  the spatial  dimension $d$, so long as it is at
least 2, the minimum Euclidean dimension that supports a magnetic flux
density.
Here $\{x^a, a = 1,2,\ldots ,d\}$ is a Cartesian coordinate system,
$\{\bm e_a, a= 1,2,\ldots ,d\}$ is  a position-independent
basis of  unit vectors, and $\delta_{ab}$
is a unimodular (determinant 1) position-independent
Euclidean metric tensor.    The unimodular
metric
defines rotational properties, and plays no role in the discussion
here, which will involve Euclidean straight-line paths, which are a 
metric-independent construction, so long as the metric tensor is not
spatially varying
(straight-line paths are minimum-length geodesics for all choices of
the Euclidean
metric tensor).
Since the metric will not be used to raise and lower spatial
indices, contravariant (upper) and covariant (lower) indices will be
distinguished, as in  $x^a$ and $\partial_a$
$\equiv$$\partial/\partial x^a$,
with implied summation over repeated upper/lower
index pairs.

The embedding of the network in Euclidean space associates each
network node with a location $\bm x_i$, and
is defined by the single-particle
operators
\begin{equation}
  \widehat X^a = \sum_{i} x^a_i \widehat P_i, 
  \quad [\widehat X^a,\widehat X^b] = [\widehat X^a_i,\widehat P_i]
  = 0.
  \label{embed}
\end{equation}
Now introduce a uniform background magnetic flux
density  in $(d\ge 2)$-dimensional Euclidean space as the gauge-invariant antisymmetric covariant tensor
\begin{equation}
F_{ab} = \partial_aA_b(\bm x) - \partial_bA_a(\bm x), \quad \text{constant},
\end{equation}
where $A_a(\bm x)$ = $\bm e_a\cdot \bm A(\bm x)$ are the covariant
components
of a gauge-dependent vector potential, the components of a
Bohm-Aharonov differential 1-form
$A_a(\bm x)dx^a$.
Then  define the modified one-particle Hamiltonian 
 $\hat h^A$ by
\begin{align}
  \langle i\sigma |\hat h^A|j\sigma'\rangle
  &= h_{i\sigma,j\sigma'} U(\bm x_i, \bm x_j) , \\
  U(\bm x_i,\bm x_j)
  &= \exp \left (i(e/\hbar)  {\textstyle\int}\bm A\cdot d\bm x\right ),
                         \nonumber \\
  {\textstyle \int} \bm A\cdot d\bm x &=
    \int_0^1d\lambda
  (x_i^a-x_j^a) A_a(\lambda \bm x_i  + (1-\lambda) \bm x_j),
  \end{align}
  where $U(\bm x_i,\bm x_j)$ is the Bohm-Aharonov factor for a
  charge-$e$ particle to tunnel along the Euclidean straight-line path
  from the embedding position $\bm x_j$ of network node $j$ to
  position $\bm x_i$  of
  network node $i$.   Note that because it is a covariant
  antisymmetric tensor, the magnetic flux density defines a
  metric-independent  Faraday 2-form $\frac{1}{2}F_{ab}dx^a\wedge
  dx^b$.
  
  The mathematical goal here is to obtain an
  explicitly-gauge-invariant form of the expansion
  \begin{equation}
    \textrm{Tr}_i\left (\widehat G^A\right )
    = \textrm{Tr}_i\left (\widehat G\right )+
    \sum_{n=1}^{\infty} \Gamma_i^{(n)}(\omega;  F_{ab}) ,
    \label{gammadef}
  \end{equation}
  where $\Gamma_i^{(n)}(\omega, \lambda F_{ab})$  =
  $\lambda^n\Gamma_i^{(n)}(\omega,F_{ab})$ ;
$\widehat G^A$ = 
  $ (\omega - \hat h^A)^{-1}$ is the network Green's function when the
  Hamiltonian is modified by the Bohm-Aharonov phase factor,
  and $\widehat G$ = $(\omega - \hat h)^{-1}$ is the Green's function
 of the network Hamiltonian $\hat h$ without this factor.
 This quantity is gauge-invariant because it can be expressed as a sum
of closed paths on the network, starting and finishing at node $i$, so the Bohm-Aharonov factor can be
expressed in a gauge-invariant way in terms of   the magnetic flux linked with the path.
It is also invariant under a translation $\widehat X^a$ $\mapsto$ $\widehat
X^a + \delta x^a\openone$ of the embedding, because the flux density
is taken to be uniform.

In a physical system, the tunneling paths will not be simple Euclidean
straight-line geodesic paths.  However, the effect of a  difference between the
true tunneling path between two orbitals
and the Euclidean straight-line path is a local effect, which does not
involve the embedding.   It merely affects the network parameters in a
gauge-invariant way.    The applied magnetic field will also in
principle locally change the shape and energy of the network orbitals.
These gauge-invariant local effects (which only depend on the magnetic
flux-density in the immediate neighborhood of  points along the path taken by the particle) will be here assumed to have  already been included as
implicit $F_{ab}$-dependence of the  network model parameters, on top
of which
the geometrical Bohm-Aharonov phase factors (which are non-local, as
they depend on the
total flux linked with the closed path of Euclidean geodesics between
network nodes) are then added.

\textbf{Statement of the result:}
The functions  $\Gamma_i^{(n)}(\omega)$ of equation (\ref{gammadef})
are  $O(\omega^{-4})$ in the large-$\omega$ limit, and can
be written as the local traces
\begin{equation}
  \Gamma_i^{(n)}(\omega)=
  \textrm{Tr}_i\left (\widehat \Gamma^{(n)}(\omega)\right
  );
  \quad \Gamma^{(n)} \equiv \sum_i \Gamma^{(n)}_i,
\end{equation}
where the
operators $\widehat \Gamma^{(n)}(\omega)$  are explicitly
gauge-invariant,
and under Hermitian
conjugation have the symmetry
\begin{equation}
\left (\widehat \Gamma^{(n)}(\omega)\right )^{\dagger} =
\widehat \Gamma^{(n)}(\omega^*).
\label{sym}
\end{equation}
The first  two of these operators can be written in terms of $\widehat
G(\omega)$ =$(\omega-\hat h)^{-1}$  to give the principal result of
this paper as
\begin{align}
  \widehat  \Gamma^{(1)}(\omega)
  &= -f_{ab}\widehat G
    \hat h_1^a \widehat G \hat h_1^b\widehat  G,
    \label{simple1}\\
  \widehat  \Gamma^{(2)} (\omega)
  &=  \tfrac{1}{2} f_{ab}f_{cd} 
    \widehat G\hat h_1^a\widehat G\hat h_2^{bc}\widehat G \hat
    h_1^d\widehat G\nonumber \\
 &\quad
                + f_{ab}f_{cd}\widehat G\hat h_1^a\widehat G\hat h_1^b\widehat G \hat
                h_1^c\widehat G \hat h_1^d \widehat G,
\label{simple2}
\\
  f_{ab}   &\equiv -\tfrac{1}{2}i(e/\hbar)F_{ab};
\end{align}
here   $\hat h_1^a$ and $\hat h_2^{ab}$ =  $\hat h_2^{ba}$  are members of  a set of embedding-dependent
gauge-invariant locally-traceless  Hermitian
operators recursively defined for $m \ge 1$ by
\begin{equation}
\hat h^a_1 = [\hat h, iX^a],\quad  h_{m+1}^{a\ldots bc} = 
 [\hat h_{m}^{a\ldots b},i\widehat X^c],
\end{equation}
where $\hat h_m^{a_1\ldots a_m}$ is fully symmetric in its
contravariant tensor indices, as  a consequence of the property
$[\widehat X^a,\widehat X^b]$ = 0.
One may also  write
\begin{align}
\hat h(\bm q) &\equiv  \widehat U(\bm q)^{-1} \hat h \widehat U(\bm q) = 
\hat h^0 + \sum_{n=1}^{\infty} \frac{1}{n!} \, \hat h_n^{a_1\ldots a_n}
q_{a_1}\ldots q_{a_n}, \nonumber \\
\widehat U(\bm q) &\equiv \exp (iq_a  \widehat X^a) =
\sum_ie^{i\bm q \cdot \bm x_i}\widehat P_i,
\end{align}
which remains well-defined even in the thermodynamic limit, when
$\widehat X^a$ itself
becomes unbounded.

For $n > 1$, the expressions for $\widehat \Gamma^{(n)}(\omega)$ are not
unique, as there are locally-traceless gauge-invariant operators  which  can be added to
to them.  For $n$ = 2, this follows from  the four linear relations
\begin{subequations}
\begin{align}
   & \textrm{Tr}_i \left (
     2\widehat \Gamma^{(2)}_{\{2,2\}}    
     + \widehat \Gamma^{(2)}_{\{2,1,1\}}    
     -\widehat \Gamma^{(2)}_{\{1,2,1\}}    
     +\widehat \Gamma^{(2)}_{\{1,1,2\}}    
    \right ) = 0 ,     \label{rela}\\
   & \textrm{Tr}_i\left ( \widehat \Gamma^{(2)}_{\{2,1,1\}}   
     -\widehat \Gamma^{(2)}_{\{1,1,2\}}    \right ) = 0, \label{relb}\\
   &\textrm{Tr}_i\left ( \widehat\Gamma^{(2)}_{\{2,1,1\}}   
     + 2\widehat\Gamma^{(2)}_{\{1,2,1\}}      
     + \widehat\Gamma^{(2)}_{\{1,1,2\}}    +
     \widehat\Gamma^{(2)}_{\{1,1,1,1\},2}   
     + \widehat \Gamma^{(2)}_{\{1,1,1,1\},3}     \right )= 0, \\
  &\textrm{Tr}_i\left (\widehat  \Gamma^{(2)}_{\{1,1,1,1\},1}                                                               
- \widehat \Gamma^{(2)}_{\{1,1,1,1\},2}    + \widehat  \Gamma^{(2)}_{\{1,1,1,1\},3}\right )    
    =  0,
    \label{reld}
 \end{align}
\end{subequations}
where
a full set of seven gauge-invariant
operators  in terms of which $\widehat \Gamma^{(2)}$
 can be represented is
 \begin{subequations}
 \begin{align}
   \widehat \Gamma^{(2)}_{\{2,2\}}
   &=  \tfrac{1}{4} f_{ac}f_{bd}\widehat G\hat
   h_2^{ab}\widehat G\hat h_2^{cd} \widehat G , \label{opa}\\
   \widehat \Gamma^{(2)}_{\{2,1,1\}}
   &=  \tfrac{1}{2} f_{ac}f_{bd}\widehat G\hat
   h_2^{ab}\widehat G \widehat h_1^c \widehat G\hat h_1^{d} \widehat G , \\
   \widehat \Gamma^{(2)}_{\{1,2,1\}}
   &=  \tfrac{1}{2} f_{ab}f_{cd}\widehat G\hat h_1^a \widehat G\hat h_2^{bc} \widehat  G\hat h_1^d
     \widehat G , \label{gammac} \\
   \widehat \Gamma^{(2)}_{\{1,1,2\}}
   &=  \tfrac{1}{2} f_{ac}f_{bd}\widehat G\hat
     h_1^a\widehat G
   h_1^b\widehat G\hat h_2^{cd} \widehat G , \\
   \widehat \Gamma^{(2)}_{\{1,1,1,1\},1}
   &=   f_{ab}f_{cd}\widehat G\hat h_1^a\widehat G\hat h_1^b
     \widehat G\hat h_1^c \widehat G\hat h_1^d \widehat G , \\
   \widehat \Gamma^{(2)}_{\{1,1,1,1\},2}
   &=   f_{ac}f_{bd}\widehat G\hat h_1^a\widehat G\hat h_1^b
     \widehat G\hat h_1^c \widehat G\hat h_1^d \widehat G , \\
   \widehat \Gamma^{(2)}_{\{1,1,1,1\},3}
   &=   f_{ad}f_{bc}\widehat G\hat h_1^a\widehat G\hat h_1^b
     \widehat G\hat h_1^c \widehat G\hat h_1^d\widehat G . \label{opd}
 \end{align}
\end{subequations}
In view of (\ref{rela})-(\ref{reld}), this basis can be said to have
dimension  $D$ = 7 with rank $R$ = 3.    The expression (\ref{simple2}) for
$\widehat \Gamma^{(2)}$ as $\widehat \Gamma^{(2)}_{\{1,2,1\}}$ +
$\widehat \Gamma^{(2)}_{\{1,1,1,1\}}$ is the unique solution to the problem of
finding the operator $\widehat \Gamma^{(2)}$  obeying
$\textrm{Tr}(\widehat \Gamma^{(2)})$ = $\Gamma^{(2)}$  as the ``most
  economical'' choice of a weighted sum of
  operators (\ref{opa})-(\ref{opd}) (\textit{i.e.}, the one
  using the smallest  number of terms, in this case 2, which is less than the rank of the basis, equal to 3).
  
  Briet \textit{et al.}\cite{briet} have studied Landau diamagnetism in the  continuum
  Schr\"{o}dinger Hamiltonian given (in units $e$ = $\hbar$ = $m_e$ =
  1) as
$\hat h$ = $\frac{1}{2}\delta^{ab}\hat p_a\hat p_b$ + $V(\bm x)$,
$\hat p_a$ $\equiv$  $-i\partial/\partial x^a$, for
which  $\hat h_1^a$ = $\delta^{ab}\hat p_a$ and $\hat h_2^{ab}$ = $\delta^{ab}\openone$.
In this model $[\hat h^a,\hat h^b]$ = 0, which is not generically
true for the network model, and $\hat h^{ab}$ is in the center of the
algebra, \textit{i.e.}, is a c-number.
For this Schr\"{o}dinger model, the authors of Ref.\onlinecite{briet} give a formula
for the quadratic term of a gauge-invariant expansion in flux density,
which they appear to  attribute to Cornean and Nenciu\cite{corneannenciu},
equivalent to
\begin{align}
  \Gamma^{(2)} &= \textrm{Tr}\left (\widehat \Gamma^{(2)}_{\{1,1,1,1\},1}
  \right )+
  \tfrac{1}{2}\delta^{ab}\delta^{cd}f_{ac}f_{bd}\textrm{
    Tr}\left ((\widehat G)^3\right ) \nonumber \\
               &\quad + f_{ac}f_{bd}\delta^{cd}
                 \textrm{Tr} \left ((\widehat G)^3\hat h_1^a\widehat G\hat h_1^b\right ) .
  \label{corn}
\end{align}
The Schr\"{o}dinger model can be considered as a continuum limit of
the generic discrete network model considered here, and the
formula (\ref{corn}) is the specialization to c-number $\hat h_2^{ab}$
$\mapsto $
$\delta^{ab}$ of an
alternative form for $\widehat \Gamma^{(2)}$
 where the  traceless operator inside the trace (\ref{rela}) has been
 added to 
(\ref{simple2}), eliminating the operator (\ref{gammac})  with label $\{1,2,1\}$, thus
giving a ``non-economical'' expression for $\widehat\Gamma^{(2)}$
with 4  (rather than 2) terms, if (\ref{sym}) is respected.

It is noteworthy that the general result reported here was obtained
purely algebraically, using simple associative-algebra methods
whch can be implemented in a finite-dimensional Hilbert space, without
requiring any consideration of a thermodynamic limit, or functional
analysis of the space of functions in continuous Euclidean space,
which  seems to be a central feature of the
Schr\"{o}dinger-differential-equation-based discussion in
Refs. \onlinecite{briet}  and \onlinecite{corneannenciu}.

For most applications, expansions beyond quadratic order become
impractical, and quadratic order is all that is needed for the
diamagnetic susceptibility.     
However, some discussion of the higher order case of general $n$ will
be given.

A generic set of  operators which can be linearly combined  to represent
$\widehat \Gamma^{(n)}$,  has the form
\begin{equation}
  \widehat\Gamma^{(n)}_{\{m_1,\ldots ,m_k\},\alpha} =\frac{f_{\bullet}\ldots
    f_{\bullet}}{m_1!\ldots m_k!}
  \widehat G \hat h^{\bullet}_{m_1}\widehat G\ldots \widehat G\hat
  h_{m_k}^{\bullet}\widehat G,
  \label{opprodx}
\end{equation}
where ``$\bullet$'' represents unspecified indices, and
$\alpha$ labels the way the contravariant indices on the $2\le k \le 2n$
instances of the symmetric tensor operators
$h_m^{a_1\ldots a_m}$ with $1 \le m \le n$, and $\sum_i m_i$ = $2n$,
are paired by
contraction with the covariant indices on
the $n$ instances of the antisymmetric tensors $f_{ab}$ 
(indices on the same instance of $\hat h_m^{\bullet}$ cannot be paired).
Given an allowed pairing, the operators (\ref{opprodx}) are defined
  up to an arbitrary choice of sign, which will here be fixed so the
  left-right order of covariant indices in $f_{ab}$ is the same is
 that of the contravariant indices in the operator product with which they are contracted.

The notation $\bm m_k$ $\equiv$ $\{m_1,\dots ,m_k\}$ will also be
used, with $\overline{\bm m}_k$= $\{m_k,\ldots ,m_1\}$ denoting the
reversed pattern.
As will be seen, the structure of $\Gamma^{(n)}$ shows  that
the expression for $\widehat \Gamma^{(n)}$ should be a weighted sum of
operators (\ref{opprodx}) with real weights:
\begin{equation}
 \widehat  \Gamma^{(n)} = \sum_{\bm m, \alpha} C_{\bm
   m,\alpha}\widehat \Gamma^{(n)}_{\bm m_k,\alpha}, \quad
 C_{\bm m_k,\alpha}  = C^*_{\bm m_k, \alpha.} .
 \label{wsum}
\end{equation}
The operators (\ref{opprodx}) have  the conjugation  property that
\begin{equation}
  \left (\widehat \Gamma^{(n)}_{\bm m_k,\alpha}(\omega)\right
  )^{\dagger} = \widehat \Gamma^{(n)}_{\overline{\bm m}_k,\bar
    \alpha}(\omega^*),
\end{equation}
which  for $\alpha$ $\ne $ $\bar \alpha$ is consistent with the chosen
sign convention for the ordering of
indices in $f_{ab}$.
Then, given that they are real,
\begin{equation}
  C_{\bm m,\alpha}= C_{\overline{\bm m}_k,\bar \alpha},
  \label{wsum1}
\end{equation}
and 
the basis set of operators (\ref{opprodx})  can be grouped into pairs
that  are each other's conjugates, or singles that
are self-conjugate.    Only a  reduced basis set that consists of the
singles, plus the sums $(\widehat \Gamma^{(n)}_{\bm m_k,\alpha} +
\widehat \Gamma^{(n)}_{\overline{\bm m}_k,\bar \alpha})$ of the
two members of each pair, is needed to represent $\widehat \Gamma^{(n)}$.

For $n$  = 2, this reduced basis set has dimension $D'$ = 6 with
rank $R'$ = 3.
For $n>2$, the number of  operators (\ref{opprodx}) grows rapidly, and
for $n=3$ a numerical linear-algebra
analysis  reveals that the basis set of  operators of type
$\widehat \Gamma^{(3)}_{\bm m_k,\alpha}(\omega)$
has $D$ =  75 with $R$ = 22, and the reduced basis set has dimension
$D'$ = 17 +  29 = 46 with  17 singles plus 29 pairs, and   rank $R'$ = 15 (so 
17 + 2$\times$29 = 75).
If an expression for $\widehat \Gamma^{(n)}$ is found as a weighted linear
combination of  a subset of $R'$ members of the reduced basis set, 
(so the set of their traces is a   linearly-independent  basis of
functions of $\omega$) it will be unique, with real weights that are
in general rational numbers
when the choice of the subset is unstructured (randomly selected).    However
 the normalization of the operators (\ref{opprodx})
 has been chosen so that, by using a suitably-chosen subset,
all the weights become  integers, and less that $R'$ of them are non-zero, as
seen in the case $n=2$, when  (\ref{simple2}) is expressed in terms
of operators (\ref{opprodx}), and in detailed investigations of
the case $n=3$ using numerical linear-algebra methods described in the Appendix.

A
principle of ``maximum economy'' suggests that  the optimal choice of
the subset of operators with linearly-independent traces is that which minimizes the number of non-zero integer
weights $C_{\bm m_k,\alpha}$  when $\widehat \Gamma^{(n)}$ is expressed in the form
(\ref{wsum}) obeying  (\ref{wsum1}), which is uniquely given once the
choice of the subset is made. Based on the case $n=2$, it is tempting
to conjecture that such an  optimal choice is unique, and provides a
``canonical'' form for $\widehat \Gamma^{(n)}$.
However, the possible uniqueness
of such a  ``most economical''  choice has
not   been checked for the case $n$ = 3, as an algorithm to search for
it was not  constructed.

In the absence of a
canonical form for $\widehat \Gamma^{(n)}$,
one generic feature
perhaps worth mentioning is that, for $n > 1$, the linear relations
between the traces always allow a form for $\widehat \Gamma^{(n)}$ to
be found that omits $\widehat \Gamma^{(n)}_{\{n,n\}}$, so  that
  $\widehat \Gamma^{(n)}(\omega)$ is $O(\omega^{-4})$ for large
  $\omega$, like its trace. While this cannot be done in the case $n=1$, the leading
$O(\omega^{-3})$ behavior of
$ \widehat \Gamma^{(1)}(\omega)$ given in (\ref{simple1}) is easily
seen to be traceless.

 \textbf{Derivation of  the result:}
The vector potential of a uniform magnetic flux density in Euclidean
space has the form
\begin{equation}
 A_a(\bm x) = -\tfrac{1}{2}F_{ab}x^b  + \partial_a\chi(\bm x) ,
\end{equation}
where $F_{ab}$is antisymmetric, and $\chi(\bm x)$ is a gauge ambiguity.
A network vector potential operator is defined by
\begin{equation}
  \widehat A_a = \sum_{i\sigma} A_a(\bm x_i)|i\sigma\rangle \langle 
  i\sigma | .
\end{equation}  
The Bohm-Aharonov phase factor is
\begin{equation}
  U(\bm x_1,\bm x_2) =
    e^{i(e/\hbar)\chi(\bm x_1)} e^{ f_{ab}x_1^ax_2^b}
    e^{-i(e/\hbar)\chi(\bm x_1)}   .
\end{equation}
The factors involving  $\chi(\bm x)$ can  be eliminated by the
gauge
transformation
\begin{equation}
  |i\sigma\rangle \mapsto e^{i(e/\hbar)\chi(\bm x_i)}|i\sigma\rangle.
\end{equation}
With this gauge choice (the ``linear gauge''), $A_a(\bm x)$ = $-\frac{1}{2}F_{ab}x^b$
is a linear function of $\bm x$, and
$U(\bm x_1,\bm x_2)$ is simply given by
$\exp (f_{ab}x_1^ax_2^b)$.
The residual gauge-dependence is in the arbitrary choice
of origin of the coordinate system: in this gauge, quantities that are
invariant under a uniform translation of the network in Euclidean
space, $\widehat X^a$ $\mapsto$ $\widehat X^a + \delta x^a\openone$,
are gauge-invariant.

In the linear gauge, the network vector potential operator is given by
\begin{equation}
  \widehat A_a = -\tfrac{1}{2}F_{ab}\widehat X^b
  \end{equation}
with $[\widehat A_a,\widehat A_b]$ =0,  $[\widehat A_a,\widehat X^b]$
= 0, and
\begin{equation}
  \widehat A_a\widehat X^a = 0.
\end{equation}
A useful property in this gauge is
\begin{equation}
  \hat h^{ab\ldots c}\widehat A_a =   \widehat A_a \hat h^{ab\ldots c} .
\label{commutex}
\end{equation}

For the change $\hat h$ $\mapsto$ $\hat h + \widehat V$,  $\widehat G$
$\mapsto$$\widehat
G^V $, which  is given by the infinite geometric series
\begin{equation}
 \widehat G^V \equiv (\omega -\hat h -\widehat V)^{-1} \equiv
  \widehat G + \widehat G\widehat V\widehat G + \widehat G\widehat V
\widehat G\widehat V\widehat G + \ldots
\end{equation}
For the change $\widehat G$ $\mapsto$  $\widehat G^A$, in the linear gauge,
\begin{align}
  \langle i\sigma | \widehat V|j\sigma'\rangle
  &=
    \langle i\sigma |\hat
    h|j\sigma \rangle
    \sum_{n=1}^{\infty}\frac{1}{n!}
    \left ( f_{ab} x^a_ix^b_j\right )^n .
\end{align}
As an operator,
\begin{align}
  \widehat V
  &= 
    \sum_{n=1}^{\infty}
    \frac{\hat h^{(n)}}{n!} ,
   \\
    \hat h^{(n)}
  &= \left( {\prod_{i=1}^ n}f_{a_ib_i}
    \right )
   \left  (\prod_{i=1}^n  \widehat X^{a_i}\right )
   \hat h
    \left   (\prod_{i=1}^n\widehat   X^{b_i}\right ).
    \label{hn}
\end{align}
Note that the substitution $\hat h$ $\mapsto$ $-\widehat G^{-1}$ can
be made in (\ref{hn}).
With the supplementary definition $\hat h^{(0)}$ $\equiv$ $\hat h$,
\begin{equation}
  f_{ab}\widehat X^a\hat h^{(n)}\widehat X^b = \hat h^{(n+1)}.
\end{equation}
Here $\hat h^{(n)}$ is a traceless Hermitian operator, which
can also be written, for  any r in the range $ [0, n]$, as
\begin{equation}
  \hat h^{(n)}
  = \left (\prod_{i=1}^r\widehat A_{a_i}\right )
 \hat h^{a_1\ldots a_n}\left (\prod_{j=r+1}^n \widehat
A_{a_j}\right ) ;
\label{hnnx}
\end{equation}
because of the property (\ref{commutex}), and mutual commutativity of
the components $\widehat A_a$,
the product of operators in (\ref{hnnx})
can be written in any order.

The series  for $\textrm{Tr}(\widehat G^A)$ in powers of $(e/\hbar)$
  has the form
 \begin{align}
   \textrm{Tr}(\widehat G^A)
   &=
     \textrm{Tr}(\widehat G)  +
     \sum_{n=1}^{\infty}
    \Gamma^{(n)},
   \\
   \sum_{n=1}^{\infty} 
    \lambda^n\Gamma^{(n)}
   &=
     \sum_{k=1}^{\infty}  \textrm{Tr}\left (\widehat G\left (\sum_{n=1}^{\infty}
\frac{1}{n!} \lambda^n\hat h^{(n)}\widehat
     G\right )^k\right ) .
     \label{gamma0}
 \end{align}
The term $\Gamma^{(n)}$ is the trace of an operator
  $\widehat G\widehat V^{(n)}\widehat G$ where
  $\widehat V^{(n)}(\omega)$  is nominally a
  polynomial of degree $n-1$ in $\widehat G$:
  \begin{align}
    \widehat V^{(n)}(\omega) &=
     \frac{1}{n!} \hat h^{(n)}  + \left (\sum_{k=1}^{n-1}
   \frac{1}{k!(n-k)!} \hat h^{(k)}\widehat G(\omega)
   \hat h^{(n-k)} \right ) +\nonumber \\
 & +   \ldots +
    \hat h^{(1)}(\widehat G(\omega)\hat h^{(1)})^{n-1}.
  \end{align}
(The nominal polynomial degree in $\widehat G$ ignores the ``hidden''
linear dependence of $\hat h^{(n)}$ on $\widehat G^{-1}$.)    
It will be useful to examine the large-$\omega$ behavior of
$\Gamma^{(n)}$.   While the operator $\widehat G\widehat
V^{(n)}\widehat G$ is $O(\omega^{-2})$ in  this limit, its trace
$\Gamma^{(n)}$ is  in fact $O(\omega^{-4})$.    The expansion is
developed using
  \begin{equation}
    \widehat G = \omega^{-1}\openone + \sum_{k=1}^{\infty}
    \omega^{-{k+1}}(\hat h)^k .
  \end{equation}
  The nominally-leading behavior of $\Gamma^{(n)}(\omega)$ at large
  $\omega$ is
  $(n!\omega^2)^{-1}\textrm{Tr}(\hat h^{(n)})$, which vanishes because
  $\hat h^{(n)}$ is traceless, so the
  apparently-leading term becomes
  \begin{equation}
    \Gamma^{(n)}(\omega) =
    \frac{1}{n!} \frac{1}{\omega^3}\textrm{Tr} \left  (\widehat O_n\right )
    + O(\omega^{-4}),
  \end{equation}
  with
  \begin{equation}
\widehat O_n =
  \hat h^{(0}\hat
        h^{(n)} + \hat
        h^{(n)}\hat h^{(0)} + \sum_{k=1}^{n-1} \frac{n!}{k!(n-k)!}\hat h^{(k)}\hat
        h^{(n-k)} .
 \end{equation}
Using (\ref{hn}), these terms can be combined to give
\begin{equation}
  \Gamma^{(n)}(\omega) =
(1-(-1)^n)\frac{1}{n!}\frac{1}{\omega^3}\textrm{Tr} \left (\hat h \hat
  h^{(n)}\right ) + O(\omega^{-4}).
\end{equation}
While (because of the prefactor)  this obviously vanishes for even $n$, it in fact vanishes for
all  $n > 0$ because  for odd $n$ $\textrm{Tr}(\hat h \hat h^{(n)})$
vanishes:
\begin{align}
  \textrm{Tr}(\hat h\hat h^{(n)})
  &= \left ( {\textstyle\prod_i}f_{a_ib_i}\right )
                                 \textrm{Tr}\left (\hat h ({\textstyle\prod_i}\widehat
  X^{a_i})\hat h ({\textstyle\prod_i}(\widehat X^{b_i})\right )
                                 \nonumber \\
  &=
                          (-1)^n\textrm{Tr}(\hat h\hat h^{(n)} ) ,
\end{align}
where the cyclic property of the trace and the antisymmetry of
$f_{ab}$ have been used.  (Note that here, wherever the cyclic
property of the trace is invoked, it always only involves moving
instances of $\widehat X^a$ between the left and  right side of the
operator-product inside the trace; since $\widehat X^a$ commutes with $\widehat
P_i$, results derived for the global trace apply
equally to the local trace.) 
Further examination shows that the $O(\omega^{-4})$ term is
generically non-vanishing, and this is the generic large-$\omega$
behavior of all terms
$\Gamma^{(n)}(\omega)$.

The explicit expressions for the first three of the $\Gamma^{(n)}$ are
\begin{align}
  \Gamma^{(1)} &= \textrm{Tr} \left (\widehat G\hat h^{(1)}\widehat G
                 \right),
                 \label{gamma00}
  \\
  \Gamma^{(2)} &= \textrm{Tr} \left (\tfrac{1}{2}\widehat G
   \hat h^{(2)}\widehat G + \widehat G\hat h^{(1)} \widehat G \hat
  h^{(1)}\widehat G\right ), \label{gamma2init}\\
  \Gamma^{(3)} &= \textrm{Tr} \left (
\tfrac{1}{6} \widehat G\hat h^{(3)} \widehat G + \tfrac{1}{2}\widehat G\hat h^{(2)}\widehat G\hat
  h^{(1)}\widehat G  + \tfrac{1}{2} \widehat G\hat h^{(1)}\widehat
                 G\hat h^{(2)}\widehat G
                 \right . \nonumber \\
  &\qquad \left . 
  +\widehat G\hat h^{(1)}\widehat G\hat h^{(1)} \widehat G \hat h^{(1)} \widehat
                 G \right ).
                 \label{gamma01}
\end{align}
Features to note are (i) these expressions are invariant under
reversing the order of the operator products inside the trace and (ii)
the operators inside the trace can be represented as  products of the
$f_{\bullet}$ times products of operators $\widehat G$, $\widehat G^{-1}$,
and $i\widehat X^{\bullet}$, added together with  with real rational
coefficients.       The manipulations to bring them to explicitly
gauge-invariant form, by using commutation relations $[\widehat
X^{\bullet},\widehat X^{\bullet}]$ = 0 and
$[h_m^{\bullet},i\widehat X^{\bullet}]$ = $\hat h_{m+1}^{\bullet}$
to eliminate all
instances of  $i\widehat X^{\bullet} $ in favor of gauge-invariant operators $\hat
h_m^{\bullet}$, together with use of the identity $\widehat
G^{-1}\widehat G$ = $\widehat G\widehat G^{-1}$ = $\openone$, can only
produce linear combinations of operators (\ref{opprodx}) with real
rational coefficients, plus traceless gauge-dependent remainder terms.

While these expressions are gauge invariant under $\widehat A_a $
$\mapsto$ $\widehat A _a  + \delta A_a\openone$, the operators
$\hat h^{(n)}$ inside
the  trace depend explicitly on $\widehat A$, and are not themselves
gauge invariant.
The difficulty with these expressions for $\Gamma^{(n)}$ is that,
while they are gauge-invariant (which is easily verified using the
numerical linear-algebra approach described in the Appendix), they are given as the
trace of a sum of operators which individually do not have
gauge-invariant
traces.

The algebraic strategy for the manipulations  to bring these
expressions to explicitly gauge-invariant form is to
reorder the gauge-dependent operator products inside the trace
to produce a sum of products of gauge-invariant operators plus a
discardable gauge-dependent
remainder term which is traceless,  so does not contribute to the function
$\Gamma^{(n)}(\omega)$.

A key  identity, is,  for any operator $\widehat O$,
\begin{equation}
[\widehat O,\widehat G] \equiv \widehat G\widehat G^{-1}\widehat O \widehat
G -  \widehat G\widehat O \widehat G^{-1} \widehat G \equiv \widehat
G[\widehat G^{-1},\widehat O]\widehat G.
\label{keyidentx}
\end{equation}
Then, in particular,
\begin{equation}
  \widehat G\hat h^a\widehat G \equiv [\widehat G,i\widehat X^a],
  \label{ghgx}
\end{equation}
which can be used to increase or decrease powers of $\widehat G$ in a
product of operators.
For example, using this one obtains
\begin{align}
  h^{(1)} &= if_{ab}\hat h^a\widehat X_b = -if_{ab}\widehat X^a\hat h^b, \label{h1}\\
  \widehat G \hat h^{(1)}\widehat G &\equiv - f_{ab}\widehat G \hat
  h^a\widehat G\hat h^b\widehat G + f_{ab}\widehat X^a\widehat G
  \widehat X^b.
  \label{gh1gx}
\end{align}

The identity that (when $n > 1$) allows $\widehat \Gamma^{(n)}_{\{n,n\}}$ to be eliminated
from expressions for $\Gamma^{(n)}$ can be derived using
(\ref{ghgx}).    First note that
\begin{align}
\widehat \Gamma^{(n)}_{\{n,n\}}    =&
   \frac{1}{n!n!} \left ({\textstyle \prod_i}f_{a_ib_i}\right ) 
    \widehat G \hat h_n^{a_1\ldots a_n}
      \widehat G \hat h_n^{b_1\ldots b_n}\widehat G \nonumber \\
  =&  \frac{1}{n!n!}\left ( {\textstyle\prod_i}
     f _{a_ib_i}\right ) 
    \widehat G [\hat h_{n-1}^{a_1\ldots a_{n-1}},i\widehat X^{a_n}]
                                  \widehat G \hat h_n^{b_1\ldots b_n}\widehat G \nonumber \\
  =&\frac{1}{n!n!}\left ({\textstyle\prod_i}f_{a_ib_i}\right ) 
    \left (-\widehat Gi\widehat X^{a_n} \hat h_{n-1}^{a_1\ldots a_{n-1}}
    \widehat G\hat h_n^{b_1\ldots b_n}\widehat G \right . \nonumber \\
    & \qquad \qquad
 +   \widehat G \hat h_{n-1}^{a_1\ldots a_{n-1}}[iX^{a_n},\widehat
        G]\hat h_n^{b_1\ldots b_n}\widehat G  \nonumber \\
    & \qquad \qquad
      \left .
    + \widehat G \hat h_{n-1}^{a_1\ldots a_n}\widehat G \hat
    h_n^{b_1\ldots b_n} i\widehat X^{a_n}\widehat G\right ) .\nonumber
\end{align}
Then, using (\ref{ghgx}), this gives the identity
\begin{equation}
 \textrm{Tr}\left (n \widehat \Gamma^{(n)}_{\{n,n\}}
  + \widehat \Gamma^{(n)}_{\{1,n-1,n\}}
   +\widehat \Gamma^{(n)}_{\{n-1,1,n\}}  -\widehat
    \Gamma^{(n)}_{\{n-1,n,1\}}\right ) = 0.
  \label{nn1}
\end{equation}
Note that this  is consistent with the property that, while
 $\widehat \Gamma^{(n)}_{\{n,n\}}(\omega)$ is $O(\omega^{-3})$ for large $
\omega$, its trace is  $O(\omega^{-4})$.   (The leading  
$O(\omega^{-3})$ term in its large-$\omega$ expansion   is
\begin{equation}
\lim_{\omega\rightarrow \infty}\widehat \Gamma^{(n)}_{\{n,n\}}(\omega)\rightarrow \left ( \prod_{i=1}^nf_{a_ib_i} \right )\frac{
  \hat h_n^{a_1\ldots a_n} \hat h_n^{b_1\ldots b_n}}{(n!)^2\omega^3} ,\nonumber
\label{traceless}
\end{equation}
which is traceless by  antisymmetry of $f_{ab}$ and the cyclic
property of the trace; tracelessness is only immediately
obvious for odd $n$, but for all $n\ge 1$ it  is easily seen after after replacement of
both instances of $h_n^{\bullet}$ by $[h_{n-1}^{\bullet},i\widehat
X^{\bullet}]$, with $``\hat h_0" \equiv \hat h$.) 
For $n$ = 2, (\ref{nn1}) corresponds to (\ref{rela}) minus
(\ref{relb}).
The other $n$ = 2 linear relations
(\ref{relb})-(\ref{reld}) can be obtained by similar manipulations,
though they were in fact initially found using
numerical linear-algebra methods.

The  explicitly-gauge-invariant form  for $\Gamma^{(n)}(\omega) $ can
be expressed in terms of the traces of 
operators of the form  (\ref{opprodx}).
However, the explicitly gauge-invariant functions defined by the traces of this
set of  operators are an overcomplete set of functions, with a high degree of
linear dependence, so there is no  unique solution to the problem of
writing $\Gamma^{(n)}$ as a sum of traces of gauge-invariant
operators.
An efficient approach to the problem, described in the Appendix, is to
use numerical linear-algebra methods to obtain a linearly-independent reduced
basis of the traces of operators (\ref{opprodx}),  then numerically solve the
linear-algebra problem of expanding $\Gamma^{(n)}(\omega)$ in this
basis,  which is a floating-point numerical  computation that produces recognizably
rational-valued expansion coefficients up to machine rounding precision.
However, to avoid just asserting  that (\ref{simple1}) and
(\ref{simple2})  ``have been shown to be true'', they will now be derived algebraically
using associative-algebra methods.

First examine $\Gamma^{(1)}$: the necessary work is already done in
(\ref{gh1gx}); since $f_{ab}\textrm{Tr}
(\widehat X^a\widehat G\widehat X^b)$ vanishes as a consequence of the
cyclic property of the trace plus antisymmetry of $f_{ab}$,
\begin{equation}
  \Gamma^{(1)} = \textrm{Tr}\left ( \widehat G \hat h^{(1)}\widehat
                 G\right )
  = - f_{ab} \textrm{Tr}\left (\widehat G\widehat h^a\widehat
                 G\widehat h^b \widehat G\right ),
\end{equation}
which  immediately confirms   (\ref{simple1}).

Next, examine $\Gamma^{(2)}$, which is the main focus of this work, as
it will be needed for a modern re-examination of formulas for Landau diamagnetism.
The starting point is the presentation (\ref{gamma2init}) of
$\Gamma^{(2)}(\omega)$ as  the gauge-invariant sum 
$\Gamma^{(2a)}$ + $\Gamma^{(2b)}$ of two terms which are not
individually gauge invariant:
\begin{align}
  \Gamma^{(2a)} & = \tfrac{1}{2}\textrm{Tr}\left (\widehat G\hat h^{(2)}
                  \widehat G \right ) ,
  \\
  \Gamma^{(2b)}  &=
                  \textrm{Tr}\left (\widehat G\hat h^{(1)}\widehat G\hat
                   h^{(1)}\widehat G\right )  \nonumber \\
  &
                   =
                 f_{ab}f_{cd}  \textrm{Tr}\left (\widehat G \hat
                 h^a\widehat X^b\widehat G\widehat X^c\hat h^d\widehat
                 G\right ),
\end{align}
where (\ref{h1}) has been used symmetrically to replace both instances
of $\hat h^{(1)}$, and begin the modification
of the second term.
Then symmetrically
use  $\widehat X^a\widehat G - \widehat G\widehat X^a$  $\equiv $ $ i\widehat
(G\hat h^a\widehat G)$,  and $\widehat X^a\hat h^b -\hat h^b\widehat
X^x$ $\equiv$ $i\hat h^{ab}$, together with
$f_{ab}f_{cd}\widehat X^b\widehat X^c$ = $f_{db}f_{ca} \widehat
X^b\widehat X^c$ and $f_{ab}\widehat X^a\hat h^b$ = $f_{ab}\hat
h^b\widehat X^a$, to get
\begin{align}
  \widehat X^b\widehat G\widehat X^c
&\equiv \widehat G\hat h^b\widehat G\hat h^c\widehat G + \tfrac{1}{2}\widehat
                                       G\hat h^{bc}\widehat G +
                                       \widehat \Delta_1^{bc} + \widehat\Delta_2^{bc}, \\
 \widehat  \Delta_1^{bc} &=  \tfrac{1}{2}\left ( \widehat G\widehat X^b\widehat
                X^c +  \widehat X^b\widehat X^c\widehat G\right
                           ), \quad
                           \\
  \widehat \Delta_2^{bc}  &=
                           \tfrac{1}{4} i\widehat G\left (
                           \{\hat h^b,\widehat X^c\} 
                         - \{\hat h^c\widehat  X^b\}
                      \right
                   ) \widehat G,
\end{align}
where $\widehat \Delta_1^{ab}$ is a symmetric tensor operator, and
$\widehat \Delta_2^{bc}$ is antisymmetric; here $\{A,B \}$ $\equiv$
$AB+BA$ is the anticommutator (symmetrized product).
  Now reassemble the full operator product in the expression for $\Gamma^{(2b)}$, and examine
\begin{align}
  \widehat \Delta_1
  &\equiv  f_{ab}f_{cd}\widehat G\hat
  h^a\widehat\Delta^{bc}_1\hat h^d\widehat G
   =
    \tfrac{1}{2}f_{ab}f_{cd}  \quad \times \nonumber \\
  &\quad\left  ( (\widehat G\hat h^a\widehat G) \widehat
    X^b\widehat X^c \hat h^d\widehat G + \widehat  G\hat h^a\widehat
    X^b\widehat X^c (\widehat G \hat h^d\widehat G)\right )
    . \nonumber
    \end{align}
    Expanding $(\widehat G\hat h^a\widehat G)$ as
    $  i(\widehat G \widehat
X^a - \widehat X^a\widehat G)$ then gives
\begin{align}
  \widehat \Delta_1 &=
    -\tfrac{1}{2}if_{ab}f_{cd} \left (\widehat X^a\widehat G \widehat
    X^b\widehat X^c
    \hat h^d\widehat G - \widehat G\hat h^a\widehat X^b\widehat
    X^c\widehat G \widehat X^d\right ) 
    \nonumber \\
  &=-\tfrac{1}{2}if_{ab}f_{cd} \left (\widehat X^a\widehat G 
    \widehat X^b\widehat X^c
    \hat h^d \widehat G - \widehat
    G\widehat h^d\widehat X^c
    X^b   \widehat G \widehat X^a \right )
    \nonumber  \\
                    &= -\tfrac{1}{4}if_{ab}f_{cd}
                      \left ([\widehat X^a,\widehat G\{\widehat X^b\widehat
    X^c,\hat h^d\}\widehat G)]\right ) \nonumber \\
                    &\qquad  -\tfrac{1}{4}if_{ab}f_{cd}\left (
                      \{\widehat X^a,\widehat G[\widehat
    X^b\widehat X^c,\hat h^d]\widehat G\}\nonumber
    \right ) .
\end{align}
Using the identities  $\textrm{Tr}( [\widehat A,\widehat B])$ $\equiv$ 0, and
$\textrm{Tr}( \{\widehat A,\widehat B\})$ $\equiv$  $2\textrm{Tr}(\widehat A\widehat
B)$, plus the cyclic property of the trace,  gives
\begin{equation}
 \textrm{Tr}\left (\widehat \Delta_1\right ) =
                                             \tfrac{1}{2} f_{ab}f_{cd}\textrm{Tr}\left
                                             ( \widehat X^a\widehat G
                                             \widehat X^c\hat
                                             h^{bd}\widehat G\right ).
\end{equation}
Next examine
\begin{align}
  \widehat \Delta_2
  &\equiv f_{ab}f_{cd} \widehat G\hat h^a
    \widehat \Delta_2^{bc}\hat h^d\widehat G 
    =\tfrac{1}{4}if_{ab}f_{cd}    \quad \times \nonumber \\
  &\quad (\widehat G\hat h^a\widehat G)
    (\{\hat h^b,\widehat X^c\} -\{\hat h^c,\widehat X^b\}
    )(\widehat G\hat h^d\widehat G) .
\end{align}
Again expanding $(\widehat G\hat h^a\widehat G)$
as     $  i(\widehat G \widehat
X^a - \widehat X^a\widehat G)$ 
gives
\begin{align}
\widehat \Delta_2  &=
    \tfrac{1}{4}if_{ab}f_{cd} \widehat G\widehat X^a(\widehat X^c\hat
    h^b - \widehat h^c\widehat X^b)\widehat X^d\widehat G \nonumber \\
  &\quad +  \tfrac{1}{4}if_{ab}f_{cd}\widehat X^a\widehat G
       (\{\hat h^b,\widehat X^c\} -\{\hat h^c,\widehat X^b\})\widehat
    G\widehat X^d\nonumber .\\
&\quad -\tfrac{1}{4}if_{ab}f_{cd}\left (
    \widehat G\widehat X^a(\{\hat h^b,\widehat X^c\}  - [\hat h^c,\widehat
    X^b])\widehat G\widehat X^d \right . \nonumber \\
  & \left . \qquad \qquad -  \widehat X^a\widehat G(\{\hat
    h^c,\widehat X^b\} - [\widehat X^c,\hat h^b])\widehat X^d\widehat
    G\right ) .\nonumber 
  \end{align}
Use $\hat h^b$ = $[\hat h, i\widehat X^b]$
to get
\begin{align}
\widehat \Delta_2
  &= -\tfrac{1}{2}f_{ab}f_{cd}\widehat G \widehat X^a\widehat X^c\hat h
    \widehat X^b\widehat X^d\widehat G \nonumber \\
  &\quad +  \tfrac{1}{4}if_{ab}f_{cd}\widehat X^a\widehat G
       (\{\hat h^b,\widehat X^c\} -\{\hat h^c,\widehat X^b\})\widehat
    G\widehat X^d\nonumber \\
  &\quad
   + \tfrac{1}{4}if_{ab}f_{cd}
    \left ( [\widehat X^a,
    \widehat G\{\hat h^c,\widehat X^b\widehat X^d\}\widehat G]\right )
    \nonumber \\
  &\quad  +\tfrac{1}{4}if_{ab}f_{cd}\left (
    \widehat G\widehat X^a[\hat h^c,\widehat X^b]\widehat G\widehat X^d
    \right . \nonumber \\
  &\quad \left .
  \qquad \qquad  - \widehat X^a \widehat G [\widehat X^c,\hat h^b]\widehat X^d\widehat
    G\right )  .
    \nonumber
  \end{align}
Now take the trace; the third term is a commutator, so is traceless, and
the second term vanishes
using the result that if $\widehat O^{\{ab\}}$ is an antisymmetric
tensor operator, $f_{ab}f_{cd}\textrm{Tr}(\widehat X^a\widehat
O^{\{bc\}}\widehat X^d)$ = 0.  The remaining two terms survive:
\begin{equation}
  \textrm{Tr}\left (\widehat \Delta_2\right ) 
  = -\tfrac{1}{2}\textrm{Tr}
    \left (\widehat G\hat h^{(2)}\widehat G\right )
    + \tfrac{1}{2}f_{ab}f_{cd}\textrm{Tr}\left ( \widehat X^a\widehat G\hat
    h^{bc}\widehat X^d\widehat G\right ) .\nonumber 
    \label{secondterm}
\end{equation}
The second  of these terms  is equal to
$-\textrm{Tr}(\widehat \Delta_1)$.
Then assembling the various parts together
gives  $\Gamma^{(2)}(\omega)$ = $\textrm{Tr}(\widehat
\Gamma^{(2)}(\omega))$,  with $\widehat \Gamma^{(2)}$ given
by  (\ref{simple2}), a representation of
$\Gamma^{(2)}$ as the trace of a  sum 
of gauge-invariant operators.   This has been achieved at the expense of a
rather onerous amount of algebraic manipulation, for which the
``roadmap'' was not obvious.  In fact, 
the result (\ref{simple2})  was initially obtained using numerical linear-algebra methods
described in the Appendix, and the
above exercise was carried out  merely to demonstrate that the result
can indeed be
obtained algebraically.

\section*{Acknowledgement}
This work was supported by the US Department of Energy,  Basic Energy
Sciences grant  DE-SC0002140.

\section*{Data availability statement}
Data sharing is not applicable to this article as no new
data were created or analyzed in this study.

\appendix
\section{Numerical linear-algebra methods}
This Appendix on numerical linear-algebra techniques adapted to this
problem is included for the possible guidance of any further explorations
of  the structure of  $\widehat \Gamma^{(n)}$ for $n \ge 3$, that
might be undertaken.

Since the process of transforming gauge-invariant (but not explicitly
gauge-invariant)  expressions for $\Gamma^{(n)}$  in
the form  of (\ref{gamma00})-(\ref{gamma01}) to the
explicitly-gauge-invariant form
(\ref{wsum}) 
purely algebraic, it can be formulated
in terms of the associative algebra of operators $\{\widehat
X^a, a=1,\ldots d\}$ which obey $[\widehat X^a,\widehat X^b]$ = 0, and
operators $\widehat G(\omega)$, $\widehat G^{-1}(\omega)$ which obey
$\widehat G\widehat G^{-1}$ = $\widehat G^{-1}\widehat G$ = $\openone$,
with no other relations, (plus a set of antisymmetric c-number 
symbols $f_{ab}$ = $-f_{ba}$).   The process can be carried using a generic
$N\times N$ matrix representation of the algebra, provided $N$ is not
so small that extra non-generic linear dependencies are introduced  (for example,
the choice $N$ = 1 is  clearly too small).

Furthermore, the algebraic structures  are independent of $\omega$, which
is the key to the numerical linear-algebra approach that uses the data provided
by multiple  instances of representations of the algebra with
different $\omega$, to define an overcomplete set of linear equations,
that can be numerically  solved using a least-squares method, to
produce (up to floating-point precision) a set of weights which are
\textit{a priori} known to be rational if computed in  exact
arithmetic, and so can be recovered from the finite-precision
complex floating-point numerical linear-algebra results, which are in
the form of simple recognizable rationals plus small residual errors
from finite machine precision.

First choose $N$ and $d \ge 2$, and  randomly populate the diagonal
entries of real diagonal $N\times N$ matrices $\widehat X^a, a =
1,\ldots, d$, and the $\frac{1}{2} d(d-1)$ independent entries of the
real  antisymmetric covariant tensor $F_{ab}$, and from it  obtain
$f_{ab}$. The algebraic structure is independent of $d$ provided
an antisymmetric symbol $f_{ab}$ exists, so the choice $d$ = 2 suffices.
The choice $N \ge 12$ worked for studying $\Gamma^{(3)}$; failures were
observed at smaller $N$, but their origin was not investigated in any detail.
Next generate a random
complex Hermitian $N\times N$ matrix $\hat h$, and diagonalize it,
to allow
$\widehat G(\omega)$  = $(\omega \openone - \hat h)^{-1}$  to be
obtained for any desired choice of  $\omega$ for which
$\widehat G(\omega)$ is non-singular.   (Since the result is algebraic,
is is not necessary to insist that $\hat h$ is Hermitian, but doing so
guarantees $\widehat G (\omega)$ is non-singular  if the 
imaginary part of $\omega$  is non-vanishing.)

Then, for a treatment of $\Gamma^{(n)}(\omega)$, compute the $N\times
  N$ matrix representations of $\hat h_m^{a_1\ldots a_m}$ for all $1
  \le m\le n$.    Next,  for any choice of the c-number covariant vector
$\delta A_a$ compute the diagonal matrices $\widehat A_a $=  $-\frac{1}{2}
F_{ab}\widehat X^b + \delta A_a\openone$, and compute the matrix representations
of $\hat h^{(1)}$, $\ldots$, $\hat h^{(n)}$.   Use these, together with
$\widehat G(\omega_i)$,  to compute $\Gamma^{(n)}(\omega_i)$,
Repeating this with different choices of $\delta A_a$ (and getting
the same result, at least to the rounding accuracy of
floating-point arithmetic) verifies the
gauge invariance of the initial expression of type
(\ref{gamma00})-(\ref{gamma01})  for $\Gamma^{(n)}(\omega)$.

Next catalog a complete set of  the 
operators $\widehat \Gamma^{(n)}_{\bm m_k,\alpha}$ (\ref{opprodx}) that can be
used to represent $\widehat \Gamma^{(n)}$.
These can be classified by partitions of $2n$ into $k$ parts equal to
or smaller than $n$, which is only  possible for $2\le k \le 2n$.
For a given partition, the set of all distinct
permutations of the
parts gives the set of  possible values of $\bm m_k$ =
$\{m_1,m_2,\ldots m_k\}$.    The operator product defined by a given
$\bm m_k$ has $2n$ contravariant indices that need to be paired by
contraction with the $2n$ covariant indices on the $n$ instances of
$f_{ab}$.  There are $(2n-1)!!$ possible pairings of $2n$ indices, but
not all are allowed, as indices on the same instance of $\hat
h^{a_!\ldots a_m}_m$ cannot be paired.  When constructing a given
operator specified by a distinct permutation of a partition and an
index pairing, the  left-right ordering of the indices in  a paring factor
$f_{ab}$ should match that  of the indices in the operator product
that it pairs.    This set of operators is invariant under a
conjugation that simultaneously reverses the order of the operator
product (and hence  the order of $\{m_1,\ldots, m_k\}$)
 and the indices of all factors $f_{ab}$, which preserves the rule for
 the index ordering in $f_{ab}$. Individual operators
 are either self-conjugate, of grouped into  mutually-conjugate
 pairs by this operation.
 The reduced basis (with dimension $D'$)  of functions  of $\omega$ 
consists of the traces
of each self-conjugate operator, plus the sum of the traces of each
mutually-conjugate pair.    For each reduced-basis function, the
operator inside the trace has the symmetry (\ref{sym}).

Denote the  set of functions just constructed by $\{f_i(\omega), i =
  1,\ldots,  D'\}$.   For $n > 1$ there is linear dependence between
  these functions, and a linearly-independent set of $R' < D'$
of  these functions must be selected.    This is done by evaluating
them on $D'$ distinct randomly-selected complex values $\{\omega_i, i =
  1,\ldots, D'\}$,    and carrying out a rank-revealing procedure
 on the $D'\times D'$ complex matrix
  $A_{ij}$ = $f_j(\omega_i)$.    Linear dependence must be removed from
  the basis set by removing redundant functions.   Once  this
  reduction has been made, there is a unique expansion of
  $\Gamma^{(n)}(\omega) $ in terms of the $R'$ retained members
  of the reduced basis, as a weighted sum with
  rational weights.    The recommended method of solution  is to use a
  least-squares solution to a substantially overcomplete set of linear equations
  \begin{equation}
    \Gamma^{(n)}(\omega_i) = \sum_{j=1}^{R'} f_j(\omega_i) C_j, \quad
    i = 1,\ldots, i_{\textrm{max}} > R',
    \label{lineq}
  \end{equation}
  where the $C_j$ will be recognizable rationals (up to floating-point
  precision).
Increasing the overcompleteness of the 
data used for the least-squares fit seems to reduce the residual errors. 
The $D'-R'$ omitted functions can also
be obtained as weighted sums (with rational weights of the $R'$
retained functions by the same method).   
Once the weights of $\Gamma^{(n)}$ and the linear relations
are obtained in rational form, they can be tested by numerical
evaluation at randomly-chosen $\omega$   All further
manipulations to optimize the expressions, or find a ``canonical''
form, are done in exact integer arithmetic.

The arbitrariness in this procedure is  in
the selection of  the linearly-independent basis of $R'$ functions out of the
full set of $D'$ functions of the reduced basis.
One
possibly-optimal structured 
approach  is to order the
basis 
by the  number of parts in $\bm m_k$ (a partial ordering), from $k=2$ to $k=2n$, and test for
linear
dependence among the traces of  the first $j$ operators in
the list, sequentially for $j$ =
$2,3,4,\ldots$.   Each time  linear dependence is found, eliminate the last
operator or operator-pair.
When the list is exhausted,
$R'$ entries will  remain.   In such a chosen subset, for $n=3$, the weights $C_{\bm
  m_k,\alpha}$ were integers, and less than $R'$ of the
basis functions were included in  the expression for $\Gamma^{(3)}$
(2 out of $R'$ = 15 integer weights  were ``unexpected'' zeroes).
A  result obtained using this method is (for real $\omega$)
\begin{align}
  \widehat \Gamma^{(3)}
  &=
    12\widehat \Gamma_{\{3,3\}}^{(14|25|36)} \nonumber \\
  &  + 12 (\widehat\Gamma_{\{3,2,1\}}^{(14|25|36)} - \widehat\Gamma_{\{1,3,2\}}^{(12|35|46)} + \textrm{H.c.})
    -28 \widehat \Gamma_{\{2,2,2\}}^{(16|23|45)} \nonumber \\
&    +12\widehat \Gamma_{\{1,2,2,1\}}^{(14|25|36)}
    -8 \widehat\Gamma_{\{1,2,2,1\}}^{(16|24|35)}
    -12\widehat \Gamma_{\{2,1,1,2\}}^{(14|25|36)} \nonumber \\
&    +8\widehat\Gamma_{\{2,1,1,2\}}^{(15|26|34)}
    -4 (\widehat\Gamma_{\{2,2,1,1\}}^{(13|25|46)} + \textrm{H.c.})
                                                                 \nonumber \\
  &
    +12(\widehat\Gamma_{\{3,1,1,1\}}^{(14|25|36)} + \textrm{H.c.})
    +4(\widehat \Gamma_{\{2,1,1,1,1\}}^{(13|26|45)}  + \textrm{H.c.})
    \nonumber \\
  & -4 \widehat \Gamma_{\{1,1,1,1,1,1\}}^{(12|34|56)}
    -6 \widehat \Gamma_{\{1,1,1,1,1,1\}}^{(16|25|34)} .
    \label{g3}
\end{align}
Here the superscripts on the operators on the RHS indicate how the six
contravariant indices (numbered from left to right)
in the operator product
should be
contracted in pairs using the  three factors of $f_{ab}$; for
example
\begin{align}
  \widehat\Gamma_{\{3,2,1\}}^{(14|25|36)}
  &= \tfrac{1}{3!2!1!}f_{a_1a_4}f_{a_2a_5}f_{a_3a_6}  \nonumber \\
  & \quad \times \widehat G \hat
h_3^{a_1a_2a_3}\widehat G\hat h^{a_4a_5}_2 
\widehat G\hat h^{a_6}_1\widehat G.
\end{align}
For complex $\omega$, the ``Hermitian conjugate''  ($\textrm{H.c.}$)
should be reinterpreted so as \textit{not} to include complex
conjugation of $\omega$ (or as Hermitian conjugation of the operator followed by
complex conjugation of $\omega$).

The integer forms of the 
linear relations (of which there are 31, for $n$ = 3) could then be used to attempt
to reduce the number of non-zero terms in the integer-weighted expression for
$\Gamma^{(n)}$, leaving its trace unchanged, but this was left undone,
in the absence of an obvious algorithm, as there was no immediate need for results with $n
> 2$.    In fact, the form (\ref{g3}) may well be optimal for $n=3$,

 \end{document}